\documentclass[a4paper,12pt]{article}
\usepackage[utf8]{inputenc}
\usepackage{cancel}
\usepackage{ulem}
\usepackage{amsfonts}
\usepackage{amssymb}
\usepackage{graphicx}
\usepackage{caption}
\usepackage[numbers,sort&compress]{natbib}
\usepackage{amsmath}
\usepackage{enumerate}
\usepackage{float}
\usepackage{mathtools}
\usepackage{subcaption}
\usepackage{color}
\usepackage{float}
\usepackage{here}

\usepackage{mathrsfs}
\usepackage{float,epsfig}
\usepackage{dcolumn}
\usepackage{authblk}
\usepackage{graphicx}
\usepackage{bm}
\usepackage{amsmath,amssymb,amsthm}
\usepackage[colorlinks=true,linkcolor=blue,citecolor=red]{hyperref}
\makeatother

\begin{document}
\title{\normalsize

{\bf \Large	Optics and Thermodynamics of Charged BTZ Black Holes with Exotic Matter Sources} }

\author[1,3]{M. A. Rbah%
\thanks{\href{mailto:Mohamedamin_rbah@um5.ac.ma}
{Mohamedamin\_rbah@um5.ac.ma}}}

\author[1,3]{S. Saoud%
\thanks{\href{mailto:soulaimane_saoud@um5.ac.ma}
{soulaimane\_saoud@um5.ac.ma}}}

\author[1,3,4]{R. Sammani%
\thanks{\href{mailto:rsammani@uottawa.ca}
{rsammani@uottawa.ca}}}

\author[1,2,3]{E. H. Saidi%
\thanks{\href{mailto:e.saidi@um5r.ac.ma}
{e.saidi@um5r.ac.ma}}}

\author[1,3]{\protect\\[0.4em]\mbox{R. Ahl Laamara}%
\thanks{\href{mailto:r.ahllaamara@um5r.ac.ma}
{r.ahllaamara@um5r.ac.ma}}}


\affil[1]{\small
LPHE-MS, Faculty of Sciences, Mohammed V University,
Rabat, Morocco}

\affil[2]{\small
Hassan II Academy of Science and Technology,
Kingdom of Morocco}

\affil[3]{\small
Centre of Physics and Mathematics, CPM-Morocco}

\affil[4]{\small
Department of Mathematics and Statistics,
University of Ottawa, Ottawa, Ontario K1N 6N5, Canada}

 \maketitle

	\begin{abstract}
{\noindent}
We study a modified charged BTZ black hole in \((2+1)\)--dimensional anti-de Sitter spacetime, where the standard geometry is surrounded by quintessence-like anisotropic matter and a cloud of strings. The corresponding metric function incorporates the effects of the electric charge and the two matter sources. The optical analysis reveals that their combined influence gives rise to a stable circular photon orbit, leading to distinctive modifications of photon dynamics and the associated energy emission. The thermodynamic analysis in a finite cavity further shows that the electric charge and exotic matter sources significantly affect the local thermal stability and phase structure of the black hole. The combined effects of these matter sources give rise to a stable photon sphere and novel thermodynamic behavior absent in the standard charged BTZ solution.

\medskip

\end{abstract}

\newpage

\tableofcontents

\section{Introduction}
\label{sec:introduction}

Three-dimensional gravity with a negative cosmological constant provides a
simple but rich framework for studying black-hole physics. The
Bañados--Teitelboim--Zanelli (BTZ) black hole is the basic example of this
class of solutions and has played an important role in lower-dimensional
gravity~\cite{Banados1992}. Although pure Einstein gravity in three dimensions
has no local propagating degrees of freedom, the BTZ spacetime still possesses
the usual black-hole properties, such as an event horizon, Hawking temperature,
and entropy. Charged and rotating generalizations of the BTZ solution provide
further examples in which the role of matter fields and conserved charges can
be studied explicitly~\cite{Martinez2000}. Since the geometry is
asymptotically anti-de Sitter, BTZ black holes also provide a useful setting
for discussions of holography and quantum gravity~\cite{Maldacena1998}.

The theory of gravity in three dimensions provides an ideal theoretical
benchmark for black-hole physics because it avoids many of the complications
encountered in higher dimensions. Since the spacetime dynamics are governed by
global and topological properties rather than gravitational waves. Moreover, the asymptotically anti-de Sitter nature of the BTZ
solution provides one of the simplest realizations of the AdS/CFT
correspondence, relating three-dimensional gravity to a two-dimensional
conformal field theory~\cite{Carlip1995,Maldacena1998,Witten1998}.

The interest in AdS backgrounds has also increased with recent developments
in the swampland program. In this context, low-energy effective theories
coupled to gravity are expected to obey consistency conditions in order to
admit a UV completion~\cite{OoguriVafa2017,Lust2019,Sammani2025BTZThreshold,10.1093/ptep/ptag126}.
Motivated by this perspective, we study matter-deformed AdS$_3$ black-hole
geometries as effective gravitational backgrounds where additional matter
sources can leave infrared signatures. Since the swampland program suggests
that not every low-energy theory coupled to gravity admits a consistent UV
completion, it is natural to ask how effective sources such as
quintessence-like anisotropic matter and clouds of strings modify the optical
and thermodynamic properties of BTZ-like black holes.

Optical observables and cavity thermodynamics provide complementary probes of
the underlying geometry. Null geodesics and circular photon orbits encode the
influence of matter sources on light propagation, whereas placing the black
hole inside a finite cavity establishes a canonical ensemble in which local
thermal equilibrium, stability, and phase behavior can be investigated in a
well-defined manner~\cite{York1986,Braden1990,Huang2022}. Together, these
approaches provide a consistent framework for understanding how additional
matter sources modify both the dynamical and thermodynamic properties of BTZ
black holes.

 In this work, we consider a charged BTZ black hole surrounded by a
quintessence-like anisotropic matter distribution and a cloud of strings. The
quintessence-like sector is described by a Kiselev-type power-law density
profile~\cite{Kiselev2003}. We use the term ``quintessence-like'' to stress
that the source is treated as an effective anisotropic fluid, rather than as a
perfect-fluid cosmological quintessence component~\cite{Visser2020}. The cloud
of strings is included following the Letelier
construction~\cite{Letelier1979}. Together with the Maxwell field, these
matter sectors deform the metric function and modify the optical and
thermodynamic properties of the black hole.

The optical sector is described by null geodesics. Circular photon orbits
encode important information about light propagation and possible optical
signatures of the spacetime. The effective-potential analysis is therefore used
to determine the existence and stability of photon
orbits~\cite{Claudel2001,Cardoso2009}. The corresponding energy emission rate
is then evaluated using the Hawking temperature and the appropriate geometric
absorption scale.

Within this framework, the cavity construction defines a canonical ensemble
through the Tolman temperature and provides a natural setting for
investigating local thermal stability and phase
transitions~\cite{York1986,Braden1990,Huang2022}. We compute the Gibbs free
energy and heat capacity to examine the effects of the electric charge,
quintessence-like matter, and the cloud of strings on the thermal phase
structure.

The present work provides a systematic analysis of the optical properties and
cavity thermodynamics of a charged BTZ black hole in the presence of
quintessence-like matter and a cloud of strings. We show that these exotic
matter sources lead to a stable photon sphere and significantly modify the
cavity phase structure, revealing features absent in the standard charged BTZ
solution.

The paper is organized as follows. In
Sec.~\ref{sec:geometrical_setup}, we present the action, the matter sources,
and the modified charged BTZ solution. In
Sec.~\ref{sec:optical_properties}, we investigate null geodesics, circular
photon orbits, and the energy emission rate. In
Sec.~\ref{sec:cavity_thermodynamics}, we analyze the cavity thermodynamics,
including the local temperature, Gibbs free energy, and heat capacity.
Finally, Sec.~\ref{sec:conclusions} summarizes the main results and discusses
possible future directions.
\section{Modified charged BTZ solution}
\label{sec:geometrical_setup}

We consider three-dimensional Einstein Gravity with a negative cosmological constant, minimally coupled to a Maxwell field, and supplemented by two effective matter sectors: a quintessence-like anisotropic component and a cloud of strings \cite{PhysRevLett.69.1849,Kiselev:2002dx,toshmatov2017rotating,deOliveira:2018weu}. The total action is given by
\begin{equation}
S
=
\frac{1}{16\pi G_3}
\int d^3x\sqrt{-g}\left(R-2\Lambda\right)
-\frac{1}{4}\int d^3x\sqrt{-g}\,
F_{\mu\nu}F^{\mu\nu}
+
\int d^3x\sqrt{-g}
\left(\mathcal{L}_{q}+\mathcal{L}_{s}\right),
\label{action_total}
\end{equation}
where the cosmological constant is fixed by \(\Lambda=-1/\ell^2\).
In what follows, we use units in which \( G_3=1\). Variation of the
action with respect to the metric gives
\begin{equation}
G_{\mu\nu}+\Lambda g_{\mu\nu}
=
T_{\mu\nu}^{(q)}
+
T_{\mu\nu}^{(s)}
+
T_{\mu\nu}^{({\rm EM})}.
\label{Einstein_equations}
\end{equation}
For the non-rotating sector, we consider a static and circularly symmetric
line element of the form
\begin{equation}
ds^2
=
-f(r)dt^2
+
\frac{dr^2}{f(r)}
+
r^2d\phi^2 .
\label{metric_ansatz}
\end{equation}

The non-rotating BTZ-type sector provides the natural setting for our analysis; the radial metric function \(f(r)\) also encodes the effects of the various matter sources.
The quintessence-like contribution is modeled as an effective anisotropic
fluid. Its nonvanishing mixed components are taken to be
\begin{equation}
T^{t}{}_{t}{}^{(q)}
=
T^{r}{}_{r}{}^{(q)}
=
-\rho_q(r),
\qquad
T^{\phi}{}_{\phi}{}^{(q)}
=
-\rho_q(r)-r\rho_q'(r),
\label{Tq}
\end{equation}
where the energy density is assumed to follow the power-law profile
\begin{equation}
\rho_q(r)
=
\alpha r^{-s},
\qquad
s=2(1+\omega_q),
\qquad
0<s<2.
\label{rhoq}
\end{equation}
Using Eq.~\eqref{rhoq}, the angular component can be written explicitly as
\begin{equation}
T^{\phi}{}_{\phi}{}^{(q)}
=
(s-1)\rho_q(r)
=
(s-1)\alpha r^{-s}.
\label{Tq_angular}
\end{equation}
This form is consistent with the conservation equation
\begin{equation}
\nabla_{\mu}T^{\mu}{}_{\nu}{}^{(q)}=0.
\label{Tq_conservation}
\end{equation}
Here, $\alpha$ controls the strength of the quintessence-like matter sector,
while the parameter $s$ determines its radial behavior and is related to the
effective equation-of-state parameter $\omega_q$ through
$s=2(1+\omega_q)$.

The cloud of strings is described by the effective stress-energy tensor
\begin{equation}
T^{t}_{\ t}{}^{(s)}
=
T^{r}_{\ r}{}^{(s)}
=
-\frac{\xi}{r},
\qquad
T^{\phi}_{\ \phi}{}^{(s)}=0 ,
\label{Ts}
\end{equation}
where \(\xi\) denotes the string-cloud density parameter. This source produces
a characteristic linear correction to the BTZ metric function \cite{Letelier:1979ej,Ghosh:2014pga,Ahmed:2025qza}.

For the electromagnetic sector, we consider a purely electric configuration,
\begin{equation}
A_\mu dx^\mu=\Phi(r)dt,
\qquad
F_{tr}=\frac{\sqrt{2}Q}{r}.
\label{electric_field}
\end{equation}

With the normalization adopted in Eq.~\eqref{electric_field}, the
nonvanishing mixed components of the electromagnetic stress-energy
tensor are
\begin{equation}
T^{t}{}_{t}{}^{({\rm EM})}
=
T^{r}{}_{r}{}^{({\rm EM})}
=
-\frac{Q^2}{r^2},
\qquad
T^{\phi}{}_{\phi}{}^{({\rm EM})}
=
\frac{Q^2}{r^2}.
\label{TEM_components}
\end{equation}
As a result of the charge normalization utilized in the present study, the
Maxwell sector produces the logarithmic contribution
\(-2Q^2\ln(r/\ell)\) in this three-dimensional charged BTZ spacetime
\cite{Martinez:1999qi,Bueno:2025dqk}.  The
non-vanishing independent components of the Einstein tensor for the line
element stated above are as follows
\begin{equation}
G^{t}_{\ t}
=
G^{r}_{\ r}
=
\frac{f'(r)}{2r},
\qquad
G^{\phi}_{\ \phi}
=
\frac{f''(r)}{2}.
\label{Einstein_tensor}
\end{equation}
The \(tt\) component of Eq.~\eqref{Einstein_equations} then reduces to the
first-order radial equation
\begin{equation}
\frac{f'(r)}{2r}
-
\frac{1}{\ell^2}
=
-\frac{\xi}{r}
-
\alpha r^{-s}
-
\frac{Q^2}{r^2}.
\label{tt_component}
\end{equation}
Equivalently,
\begin{equation}
f'(r)
=
\frac{2r}{\ell^2}
-
2\xi
-
2\alpha r^{1-s}
-
\frac{2Q^2}{r}.
\label{f_derivative}
\end{equation}
For \(s\neq2\), direct integration gives the generalized charged BTZ metric
function
\begin{equation}
f(r)
=
-M
+
\frac{r^2}{\ell^2}
-
2\xi r
-
\frac{2\alpha}{2-s}r^{2-s}
-
2Q^2\ln\left(\frac{r}{\ell}\right),
\label{f_final_mass}
\end{equation}
where \(M\) comes from the integration constant and is identified with the
mass parameter of the black hole. Each term of the metric function has a clear
geometrical or physical interpretation \cite{Chen:2025nto}. The quadratic term is due to the AdS background, the linear term results from the string cloud,
the power-law correction arises from anisotropic matter sensitive to
quintessence, and the logarithmic term is associated with the electric
contribution expected from charged BTZ geometries.

The outer horizon is determined by the largest positive root of
\begin{equation}
f(r_+)=0.
\end{equation}
Solving this condition for the mass parameter yields
\begin{equation}
M
=
\frac{r_+^2}{\ell^2}
-
2\xi r_+
-
\frac{2\alpha}{2-s}r_+^{2-s}
-
2Q^2\ln\left(\frac{r_+}{\ell}\right).
\label{mass_horizon_relation}
\end{equation}
 Substituting
Eq.~\eqref{mass_horizon_relation} into Eq.~\eqref{f_final_mass}, one obtains
the horizon-normalized form
\begin{equation}
f(r)
=
\frac{r^2-r_+^2}{\ell^2}
-
2\xi(r-r_+)
-
\frac{2\alpha}{2-s}
\left(
r^{2-s}-r_+^{2-s}
\right)
-
2Q^2\ln\left(\frac{r}{r_+}\right).
\label{f_final_horizon}
\end{equation}
This expression makes the horizon structure explicit and automatically
satisfies \(f(r_+)=0\).

Finally, the standard non-rotating BTZ black hole is recovered by switching
off all additional matter contributions,
\begin{equation}
\alpha=\xi=Q=0.
\label{BTZ_limit_condition}
\end{equation}
In this limit, Eq.~\eqref{f_final_mass} reduces to
\begin{equation}
f(r)
=
-M+\frac{r^2}{\ell^2},
\label{BTZ_limit}
\end{equation}
confirming that \(\alpha\), \(\xi\), and \(Q\) respectively parametrize the
deformations induced by the quintessence-like sector, the cloud of strings,
and the electric charge.

\section{Photon Sphere and Energy Emission Rate}
\label{sec:optical_properties}
\subsection{Optical Setup and Null Geodesics}
\label{subsec:optical_setup}

In asymptotically AdS$_3$ spacetimes, the notion of a physical observer at spatial infinity is not operational in the same way as in asymptotically flat geometries. Therefore, to define optical observables in a well-controlled manner, we introduce a
cavity wall at \(r=r_{\rm cav}\). The exterior region relevant for the cavity
ensemble is then restricted to
\begin{equation}
r_+ < r < r_{\rm cav}.
\label{eq:cavity_domain}
\end{equation}
In this setup, all locally measured quantities are defined with respect to a
static observer located at the cavity wall \cite{Huang:2021eby}.

We consider the static and circularly symmetric line element defined in
Eq.~\eqref{metric_ansatz} and study the propagation of massless particles \cite{Martinez:2019nor,Chen:2025nto}.
The null geodesics follow from the Lagrangian
\begin{equation}
\mathcal{L}
=\frac{1}{2}\left[-f(r)\dot t^{\,2}+\frac{\dot r^{\,2}}{f(r)}+r^2\dot\phi^{\,2}\right]=0,
\label{eq:null_lagrangian}
\end{equation}
where the dot denotes differentiation with respect to an affine parameter $\lambda$.
Since the metric is independent of $t$ and $\phi$, there exist two conserved quantities, interpreted as the photon energy and angular momentum,
\begin{equation}
E=f(r)\dot t,
\qquad
L=r^2\dot\phi.
\label{eq:conserved_EL}
\end{equation}
Substituting these relations into the null constraint $\mathcal{L}=0$ yields the radial equation
\begin{equation}
\dot r^{\,2}+V_{\rm eff}(r)=E^2,
\qquad
V_{\rm eff}(r)=\frac{L^2}{r^2}\,f(r).
\label{eq:radial_equation_veff}
\end{equation}
As a result, the structure of the optical space of the geometry is encoded in the effective potential \(V_{\rm eff}\), which determines both the existence
of circular photon orbits, as well as their type if they do exist.

In order for a circular null orbit, namely the photon sphere, to exist, it must
meet both conditions defined by the radial velocity being zero and by the radial
derivative of the effective potential being equal to zero,
\begin{equation}
\dot r=0,
\qquad
\frac{dV_{\rm eff}}{dr}=0.
\label{eq:photon_sphere_conditions}
\end{equation}
These two conditions can be expressed in a single compact equation involving
both the metric function and its radial derivative,
\begin{equation}
r\,f'(r)-2f(r)=0,
\label{eq:photon_sphere_rf}
\end{equation}
and the radius of the photon sphere can then be solved for as the quantity
\(r=r_{\rm ph}\) which meets these two conditions and is located outside of the
event horizon,
\begin{equation}
r_{\rm ph}>r_+.
\label{eq:photon_sphere_outside}
\end{equation}
It should also be noted that, in the absence of an unstable circular null orbit
outside the event horizon, the standard black-hole shadow construction is no
longer applicable, since the shadow boundary is conventionally determined by
the unstable photon sphere~\cite{Claudel2001,Cardoso2009,Perlick2022}. Rather, the energy
emission rate can be computed from the Hawking temperature and an effective
absorption cross section based on the horizon radius, as discussed in
Sec.~\ref{subsec:emission_rate}.
\subsection{Analytical circular null orbits for \texorpdfstring{$s=1$}{s=1}}
\label{subsec:photon_sphere_s1}

For \(s=1\), the quintessence contribution has the same radial dependence as
the string-cloud term. It is therefore convenient to introduce
\begin{equation}
\beta \equiv \alpha+\xi .
\label{eq:beta_s1}
\end{equation}
The metric function can then be
expressed in horizon-normalized form in terms of the outer horizon radius.
\begin{equation}
f(r)
=
\frac{r^2-r_+^2}{\ell^2}
-
2\beta(r-r_+)
-
2Q^2\ln\left(\frac{r}{r_+}\right),
\label{eq:f_s1_photon}
\end{equation}
which automatically satisfies \(f(r_+)=0\).

For null geodesics, the radial equation is
\begin{equation}
\dot r^{\,2}+V_{\rm eff}(r)=E^2,
\qquad
V_{\rm eff}(r)=\frac{L^2}{r^2}f(r),
\label{eq:Veff_s1}
\end{equation}
where \(E\) and \(L\) are the conserved photon energy and angular momentum.

Circular null orbits are determined by
\begin{equation}
\frac{dV_{\rm eff}}{dr}=0,
\qquad
\Longleftrightarrow
\qquad
r f'(r)-2f(r)=0 .
\label{eq:photon_condition_s1}
\end{equation}
For Eq.~\eqref{eq:f_s1_photon}, one has
\begin{equation}
f'(r)
=
\frac{2r}{\ell^2}
-
2\beta
-
\frac{2Q^2}{r}.
\label{eq:fprime_s1_photon}
\end{equation}
Substitution into Eq.~\eqref{eq:photon_condition_s1} gives
\begin{equation}
\beta r_{\rm ph}
+
2Q^2\ln\left(\frac{r_{\rm ph}}{r_+}\right)
=
2\beta r_+
+
Q^2
-
\frac{r_+^2}{\ell^2}.
\label{eq:rph_transcendental}
\end{equation}
The logarithmic charge term makes this equation transcendental.

For \(Q\neq0\) and \(\beta\neq0\), Eq.~\eqref{eq:rph_transcendental} can be
solved in terms of the Lambert function. Defining
\begin{equation}
z
=
\frac{\beta r_+}{2Q^2}
\exp\left[
\frac{
2\beta r_+
+
Q^2
-
r_+^2/\ell^2
}{
2Q^2
}
\right],
\label{eq:z_lambert_s1}
\end{equation}
one obtains
\begin{equation}
r_{\rm ph}^{(k)}
=
\frac{2Q^2}{\beta}
W_k(z),
\label{eq:rph_lambert_s1}
\end{equation}
where \(W_k\) is the \(k\)-th branch of the Lambert function. Since
\(\beta>0\) in the parameter sector analyzed here, \(z>0\), and the relevant
real solution is provided by the principal branch \(W_0\). The root must be
selected outside the horizon \cite{Corless:1996LambertW,BravoGaete:2019rci}. For optical applications based on a real impact
parameter, it should also lie in the static exterior region, since
\begin{equation}
b_{\rm ph}^2
=
\frac{L^2}{E^2}
=
\frac{r_{\rm ph}^2}{f(r_{\rm ph})}.
\label{eq:bph_s1}
\end{equation}

The stability of the circular null orbit is determined by the second
derivative of the effective potential. In general
\begin{equation}
V_{\rm eff}''(r)
=
L^2
\left[
\frac{f''(r)}{r^2}
-
\frac{4f'(r)}{r^3}
+
\frac{6f(r)}{r^4}
\right].
\label{eq:Veff_second_general_s1}
\end{equation}
Using the circular-orbit condition
\(r_{\rm ph}f'(r_{\rm ph})=2f(r_{\rm ph})\), this reduces to
\begin{equation}
V_{\rm eff}''(r_{\rm ph})
=
\frac{L^2}{r_{\rm ph}^4}
\left[
r_{\rm ph}^2 f''(r_{\rm ph})
-
2f(r_{\rm ph})
\right].
\label{eq:Veff_second_reduced_s1}
\end{equation}
For the metric \eqref{eq:f_s1_photon},
\begin{equation}
f''(r)
=
\frac{2}{\ell^2}
+
\frac{2Q^2}{r^2}.
\label{eq:fsecond_s1}
\end{equation}
Moreover, Eq.~\eqref{eq:rph_transcendental} gives
\begin{equation}
f(r_{\rm ph})
=
\frac{r_{\rm ph}^2}{\ell^2}
-
\beta r_{\rm ph}
-
Q^2 .
\label{eq:f_ph_s1_compact}
\end{equation}
Hence
\begin{equation}
V_{\rm eff}''(r_{\rm ph})
=
\frac{2L^2}{r_{\rm ph}^4}
\left(
\beta r_{\rm ph}+2Q^2
\right).
\label{eq:Veff_second_final_s1}
\end{equation}
Since \(\beta>0\), \(Q^2\geq0\), and \(r_{\rm ph}>0\), any admissible circular
null orbit in this sector satisfies
\begin{equation}
V_{\rm eff}''(r_{\rm ph})>0 .
\label{eq:positive_beta_stability_s1}
\end{equation}
Thus, in the positive-density sector, circular null orbits correspond to
stable null trapping rather than to unstable light rings. The alternative
case \(\beta<0\) would require an effective negative matter contribution, and is not part of the physical parameter space considered
in this work.

The uncharged case follows by setting \(Q=0\) directly in
Eq.~\eqref{eq:rph_transcendental}. One obtains
\begin{equation}
r_{\rm ph}^{(Q=0)}
=
2r_+
-
\frac{r_+^2}{\beta\ell^2},
\label{eq:rph_uncharged_s1}
\end{equation}
with
\begin{equation}
V_{\rm eff}''(r_{\rm ph}^{(Q=0)})
=
\frac{2L^2\beta}{\left(r_{\rm ph}^{(Q=0)}\right)^3}.
\label{eq:uncharged_stability_s1}
\end{equation}
Therefore, the uncharged circular orbit, whenever it lies in the exterior
region, is stable in the positive-\(\beta\) sector.

For completeness, the formal limiting case \(\beta=0\) gives
\begin{equation}
r_{\rm ph}^{(\beta=0)}
=
r_+
\exp\left[
\frac{1}{2}
-
\frac{r_+^2}{2Q^2\ell^2}
\right],
\label{eq:rph_beta_zero_s1}
\end{equation}
and
\begin{equation}
V_{\rm eff}''(r_{\rm ph}^{(\beta=0)})
=
\frac{4L^2Q^2}
{\left(r_{\rm ph}^{(\beta=0)}\right)^4}
>0 .
\label{eq:beta_zero_stability_s1}
\end{equation}
When a charged \(\beta=0\) circular orbit is admissible, it is also stable.

Table~\ref{tab:rph_general_s_values_corrected} shows that the formal
roots of the circular null-orbit equation lie inside the outer horizon
for all the representative configurations considered. More precisely,
the complete horizon analysis shows that these roots are located
between the inner and outer horizons, where
$f(r_{\rm ph})<0$ and
$b_{\rm ph}^{2}=r_{\rm ph}^{2}/f(r_{\rm ph})<0$.
Consequently, none of these roots represents a physical circular null
orbit in the static exterior region.

\begin{table}[H]
\centering
\caption{ Recalculated roots of the circular null-orbit equation
$r f'(r)-2f(r)=0$. Here, $r_+$ denotes the largest positive root of
$f(r)=0$. The reported values of $r_{\rm ph}$ are formal mathematical
roots; in all cases, they satisfy $r_{\rm ph}<r_+$ and therefore do not
represent physical circular null orbits in the exterior region.
}
\label{tab:rph_general_s_values_corrected}
\begin{tabular}{ccccccc}
\hline
\(\alpha\) & \(\xi\) & \(s\) & \(Q\) & \(r_+\) & \(\ell\) & \(r_{\rm ph}\) \\
\hline
\(0.02\) & \(0.02\) & \(0.50\) & \(0.50\) & \(0.500000\) & \(0.50\) & \(0.117709\) \\
\(0.05\) & \(0.05\) & \(1.00\) & \(0.50\) & \(0.606532\) & \(1.00\) & \(0.547374\) \\
\(0.05\) & \(0.05\) & \(1.25\) & \(0.50\) & \(0.615479\) & \(1.00\) & \(0.550876\) \\
\(0.10\) & \(0.10\) & \(1.00\) & \(0.50\) & \(0.725434\) & \(1.00\) & \(0.589277\) \\
\(0.20\) & \(0.10\) & \(1.50\) & \(0.60\) & \(1.101088\) & \(1.00\) & \(0.674452\) \\
\hline
\end{tabular}
\end{table}

The numerical values show that the circular-orbit radius is sensitive to the
charge, the matter-sector parameters and the exponent \(s\). They complement
the closed-form Lambert-\(W\) result obtained in the special \(s=1\) case.
\subsection{Energy emission rate}
\label{subsec:emission_rate}

The energy emission rate gives another way to characterize the optical and
thermal response of the geometry. In the geometric-optics regime, the spectrum
may be modeled by combining the Hawking temperature with an effective
absorption scale \cite{Wei:2011zw,Decanini:2011xi}. In the positive matter
sector considered here, no standard unstable photon-sphere separatrix appears
outside the horizon. We therefore take the dominant absorption scale to be set
by the horizon circumference,
\begin{equation}
\sigma_{\rm abs}\simeq 2\pi r_+ .
\label{eq:sigma_abs_horizon}
\end{equation}
The corresponding emission spectrum is written as
\begin{equation}
\frac{d^2E}{dt\,d\bar{\omega}}
=
\frac{4\pi}{3}\,
\sigma_{\rm abs}\,
\frac{\bar{\omega}^2}
{\exp\!\left(\bar{\omega}/T_H\right)-1},
\label{eq:emission_general}
\end{equation}
where \(\bar{\omega}\) is the emitted frequency and \(T_H\) is the Hawking
temperature \cite{Hyun:1994na}. Using Eq.~\eqref{eq:sigma_abs_horizon}, one obtains
\begin{equation}
\frac{d^2E}{dt\,d\bar{\omega}}
\simeq
\frac{8\pi^2}{3}\,
r_+\,
\frac{\bar{\omega}^2}
{\exp\!\left(\bar{\omega}/T_H\right)-1}.
\label{eq:emission_final}
\end{equation}
The Hawking temperature is determined by the surface gravity,
\begin{equation}
T_H=\frac{f'(r_+)}{4\pi}.
\label{eq:TH_surface_gravity}
\end{equation}
Thus, the emission rate is governed by two geometric quantities: the horizon
radius \(r_+\), which sets the effective absorption length, and the surface
gravity, which determines the thermal factor.

\begin{figure}[t]
    \centering
    \begin{subfigure}{0.48\textwidth}
        \centering
        \includegraphics[width=\textwidth]{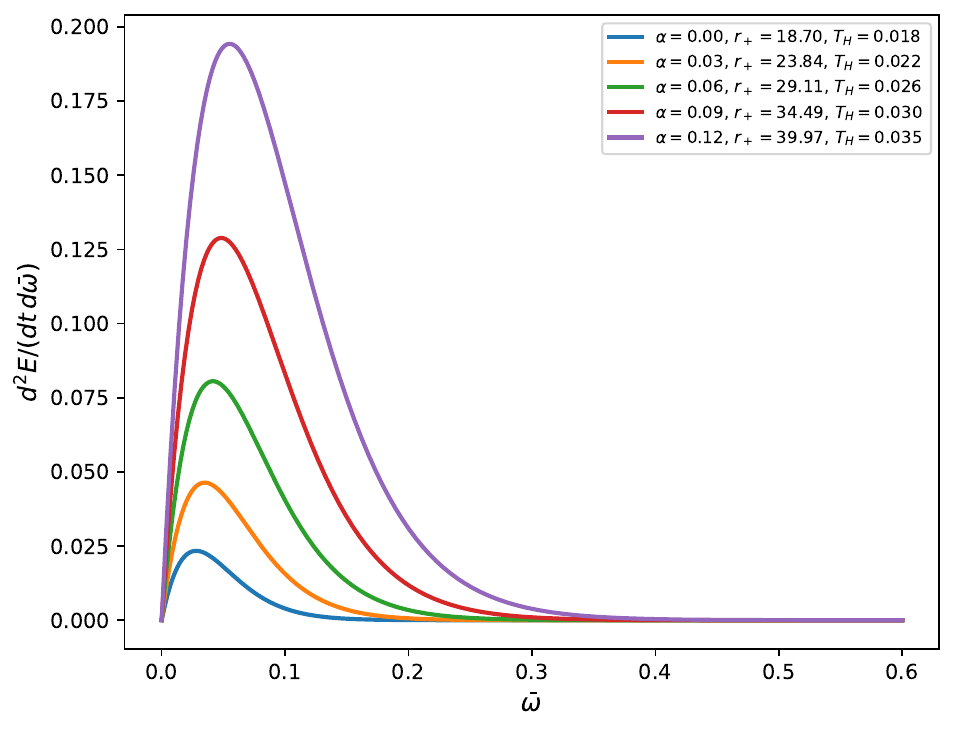}
        \caption{Variation with \(\alpha\) at fixed \(\xi\).}
        \label{fig:emission_alpha}
    \end{subfigure}
    \hfill
    \begin{subfigure}{0.48\textwidth}
        \centering
        \includegraphics[width=\textwidth]{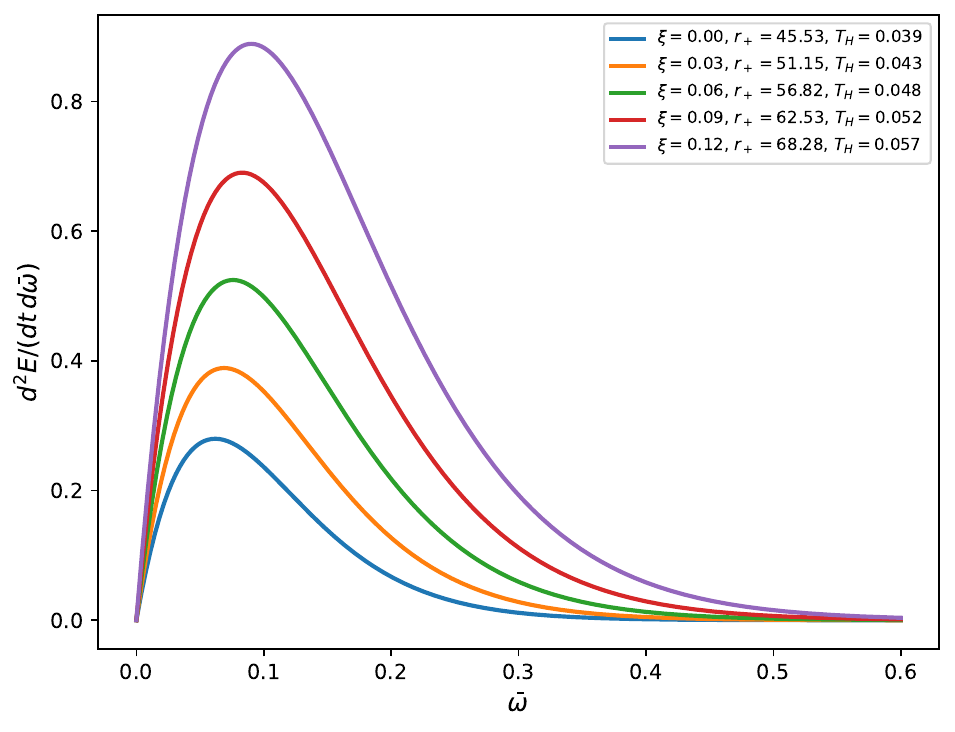}
        \caption{Variation with \(\xi\) at fixed \(\alpha\).}
        \label{fig:emission_xi}
    \end{subfigure}
    \caption{Energy emission rate as a function of the frequency
    \(\bar{\omega}\) for different matter-sector parameters. Only
    configurations with \(T_H>0\) are shown.}
    \label{fig:emission_rate_subfigures}
\end{figure}

Figure~\ref{fig:emission_rate_subfigures} shows that the surrounding matter
sector strongly affects the radiated flux. Increasing the quintessence-like
parameter \(\alpha\) enhances the emission peak and shifts the dominant
emission toward larger frequencies. This follows from
Eq.~\eqref{eq:emission_final} in the displayed parameter range, increasing
\(\alpha\) increases both the horizon radius and the Hawking temperature.
The larger \(r_+\) enhances the effective absorption length, while the larger
\(T_H\) weakens the exponential suppression in the Planck factor 

A similar effect is obtained when the string-cloud parameter \(\xi\) is
increased. The string cloud modifies the lapse function, shifts the outer
horizon, and changes the surface gravity. For the configurations shown in
Fig.~\ref{fig:emission_xi}, both \(r_+\) and \(T_H\) increase with \(\xi\),
leading to a larger emission amplitude. In both panels, the spectrum has the
usual thermal profile; it rises from small values at low frequency, reaches a
maximum, and is exponentially suppressed at large \(\bar{\omega}\). The
matter fields, therefore, affect the emission rate indirectly, through their
effect on the horizon scale and Hawking temperature.
\section{Cavity Thermodynamics and Thermal Stability}
\label{sec:cavity_thermodynamics}
The thermodynamic behavior of black holes in asymptotically AdS spacetimes is
sensitive to the choice of ensemble and to the location of the observer
\cite{York1986,Braden1990}. In this section, we study the thermal properties
of the generalized charged BTZ black hole by placing the system inside a finite
cavity. This construction allows to define local thermodynamic quantities
at a finite radial boundary and to analyze how the quintessence-like matter
field, the cloud of strings and the electric charge modify the temperature,
free energy and stability of the black-hole branches \cite{Huang2022}. We
first derive the cavity temperature through the Tolman redshift relation, and
then express the Hawking temperature and the free energy as functions of the
entropy in order to investigate the possible thermal phase structure
\cite{Banados1992,Huang2022}.
\subsection{Local Thermodynamics in a Finite Cavity}
\label{subsec:local_cavity_thermodynamics}
We now study the thermodynamics of the generalized charged BTZ black hole inside a finite cavity \cite{Huang2022,Huang:2022agr}. 
Using the metric function \ref{f_final_horizon}
the horizon condition allows one to write
\begin{equation}
f(r)
=
\frac{r^2-r_+^2}{\ell^2}
-
2\xi(r-r_+)
-
\frac{2\alpha}{2-s}
\left(
r^{2-s}-r_+^{2-s}
\right)
-
2Q^2\ln\left(\frac{r}{r_+}\right).
\label{eq:f_cavity_horizon}
\end{equation}
This form automatically satisfies $f(r_+)=0$ and is convenient for the cavity analysis.

The Hawking temperature is obtained from the surface gravity,
\begin{equation}
T_H
=
\frac{f'(r_+)}{4\pi}
=
\frac{1}{2\pi}
\left(
\frac{r_+}{\ell^2}
-
\xi
-
\alpha r_+^{1-s}
-
\frac{Q^2}{r_+}
\right).
\label{eq:TH_cavity}
\end{equation}
The local temperature measured by an observer at the cavity wall follows from the Tolman redshift relation,
\begin{equation}
T_{\rm cav}
=
\frac{T_H}{\sqrt{f(r_{\rm cav})}}.
\label{eq:Tcav_def}
\end{equation}
Therefore,
\begin{equation}
T_{\rm cav}
=
\frac{
\displaystyle
\frac{1}{2\pi}
\left(
\frac{r_+}{\ell^2}
-
\xi
-
\alpha r_+^{1-s}
-
\frac{Q^2}{r_+}
\right)
}{
\displaystyle
\sqrt{
\frac{r_{\rm cav}^2-r_+^2}{\ell^2}
-
2\xi(r_{\rm cav}-r_+)
-
\frac{2\alpha}{2-s}
\left(
r_{\rm cav}^{2-s}-r_+^{2-s}
\right)
-
2Q^2\ln\left(\frac{r_{\rm cav}}{r_+}\right)
}
}.
\label{eq:Tcav_general}
\end{equation}
The condition $f(r_{\rm cav})>0$ ensures that the cavity wall lies in the static region outside the horizon.

For the special case $s=1$, the metric function becomes
\begin{equation}
f(r)
=
\frac{r^2-r_+^2}{\ell^2}
-
2\beta(r-r_+)
-
2Q^2\ln\left(\frac{r}{r_+}\right),
\label{eq:f_cavity_s1}
\end{equation}
and the local cavity temperature reduces to
\begin{equation}
T_{\rm cav}
=
\frac{
\displaystyle
\frac{1}{2\pi}
\left(
\frac{r_+}{\ell^2}
-
\beta
-
\frac{Q^2}{r_+}
\right)
}{
\displaystyle
\sqrt{
\frac{r_{\rm cav}^2-r_+^2}{\ell^2}
-
2\beta(r_{\rm cav}-r_+)
-
2Q^2\ln\left(\frac{r_{\rm cav}}{r_+}\right)
}
}.
\label{eq:Tcav_s1}
\end{equation}

   \begin{figure}[H]
    \centering
    \begin{subfigure}{0.48\textwidth}
        \centering
        \includegraphics[width=\textwidth]{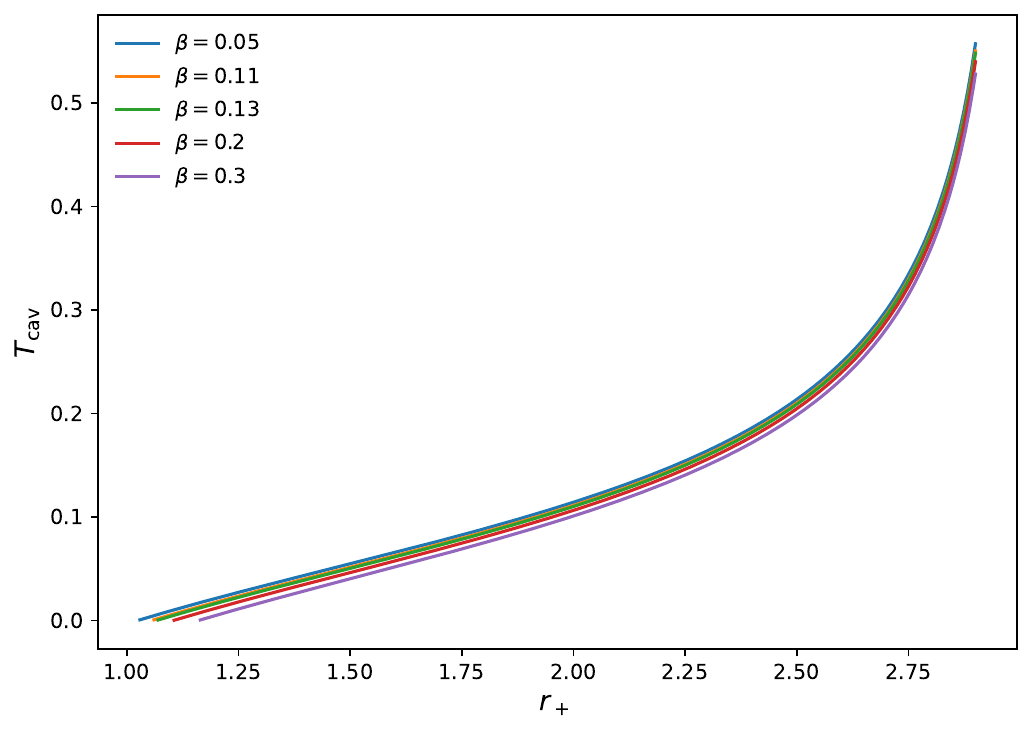}
        \subcaption{$Q =1$}
        \label{fig:temp_Q}
    \end{subfigure}
    \hfill
    \begin{subfigure}{0.48\textwidth}
        \centering
        \includegraphics[width=\textwidth]{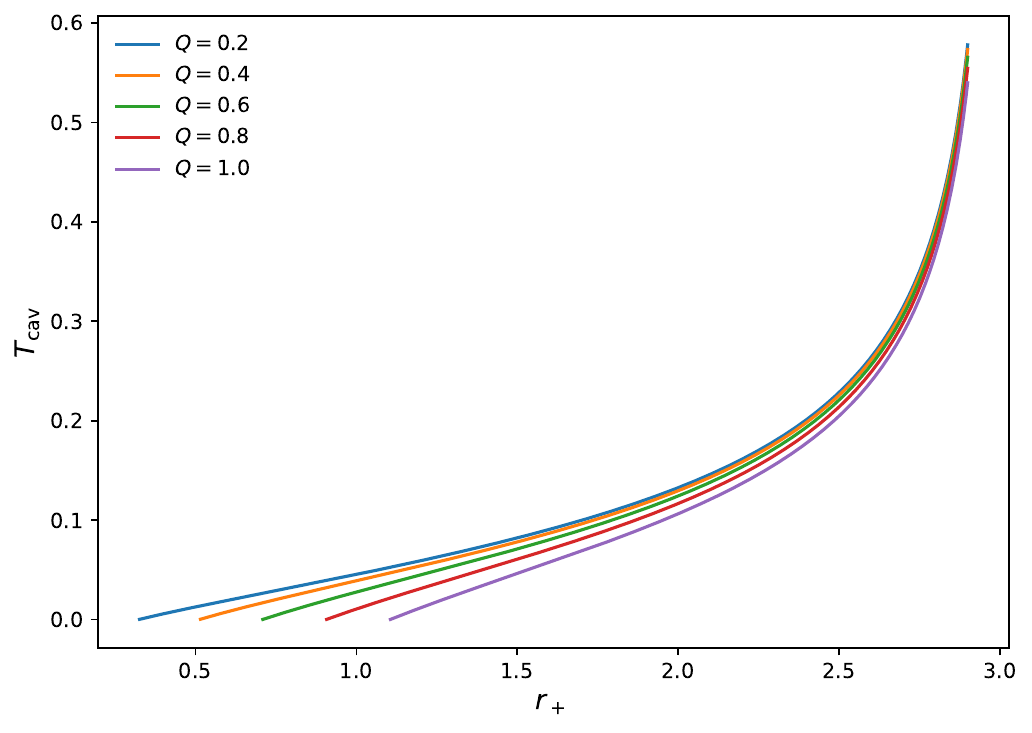}
        \subcaption{$\beta=0.2$}
        \label{fig:temp_beta}
    \end{subfigure}
    \caption{Cavity temperature for different $r_{\text{cav}} = 3.0$, $\ell = 1.0$.} 
    \label{fig:temperatures_cavite}
\end{figure}
Figure~\ref{fig:temperatures_cavite} represents the temperature as a function of the horizon radius for a
fixed cavity size \(r_{\rm cav}=3.0\) and AdS scale \(\ell=1.0\). In the left
plot, the electric charge is kept constant while the effective matter parameter
\(\beta\) is varied. This isolates the effects of the quintessence-like sector
and the string cloud on the local thermal behavior. In the right plot, the
effective matter parameter is kept constant while the electric charge \(Q\) is
varied, showing the effect of the electromagnetic sector on both the Hawking
temperature and the Tolman redshift factor. The extrema of the local cavity
temperature indicate possible changes of thermodynamic stability between
different black-hole branches.
\subsection{Entropy Representation and Thermal Stability}
\label{subsec:entropy_thermal_stability}

To make the phase structure more transparent, we use the entropy as the
thermodynamic variable. In \((2+1)\) dimensions, the black-hole entropy is proportional to the horizon circumference
rather than to a horizon area
\cite{Mendez-Zavaleta:2026rgg,Ladghami:2024qaf,Kumar:2024pkh,Bhattacharjee:2024sfe}. It is given by
\begin{equation}
S=\frac{\pi r_+}{2},
\qquad
r_+=\frac{2S}{\pi}.
\label{eq:entropy_rplus}
\end{equation}
Substituting this relation into the Hawking temperature, one obtains
\begin{equation}
T_H(S,Q,P,\alpha,\xi)
=
\frac{1}{2\pi}
\left[
\frac{2S}{\pi\ell^2}
-\xi
-\alpha\left(\frac{2S}{\pi}\right)^{1-s}
-\frac{\pi Q^2}{2S}
\right].
\label{eq:TH_S_general}
\end{equation}
Equivalently, using the thermodynamic pressure
\begin{equation}
P=\frac{1}{8\pi\ell^2},
\label{eq:pressure_ads}
\end{equation}
the temperature can be written as
\begin{equation}
T_H(S,Q,P,\alpha,\xi)
=
\frac{8PS}{\pi}
-
\frac{\xi}{2\pi}
-
\frac{\alpha}{2\pi}
\left(\frac{2S}{\pi}\right)^{1-s}
-
\frac{Q^2}{4S}.
\label{eq:TH_S_pressure}
\end{equation}

The mass parameter expressed as a function of the entropy is
\begin{equation}
M(S,Q,P,\alpha,\xi)
=
\frac{4S^2}{\pi^2\ell^2}
-
\frac{4\xi S}{\pi}
-
\frac{2\alpha}{2-s}
\left(\frac{2S}{\pi}\right)^{2-s}
-
2Q^2\ln\left(\frac{2S}{\pi\ell}\right).
\label{eq:M_S}
\end{equation}
In the fixed-charge ensemble, the relevant free energy is defined by
\begin{equation}
G(S,Q,P,\alpha,\xi)=M-T_HS.
\label{eq:G_def}
\end{equation}
Using Eqs.~\eqref{eq:TH_S_general} and \eqref{eq:M_S}, we obtain

\begin{equation}
G(S,Q,P,\alpha,\xi)
=
\frac{24PS^2}{\pi}
-
\frac{7\xi S}{2\pi}
-
\frac{\alpha(s+6)}{4(2-s)}
\left(\frac{2S}{\pi}\right)^{2-s}
-
2Q^2\ln\left(
\frac{2S}{\pi}\sqrt{8\pi P}
\right)
+
\frac{Q^2}{4}.
\label{eq:G_S_pressure}
\end{equation}

The local thermodynamic stability is characterized by the heat capacity,
defined in the fixed-charge ensemble as
\begin{equation}
C
=
T_H\left(\frac{\partial S}{\partial T_H}\right)_{Q,P,\alpha,\xi} .
\label{eq:heat_capacity_short}
\end{equation}
A positive heat capacity corresponds to a locally stable black-hole branch,
whereas a negative heat capacity signals thermodynamic instability.
Moreover, the divergences of $C$ occur at the extrema of $T_H(S,Q,P,\alpha,\xi)$ and
therefore indicate possible transitions between different thermal branches.
For a clearer comparison near the critical point, we use the reduced variables
$\tilde{S}=S/S_c$ and $\tilde{C}=C/C_c$.

\begin{figure}[H]
\centering

\begin{subfigure}{0.32\textwidth}
\centering
\includegraphics[width=\linewidth]{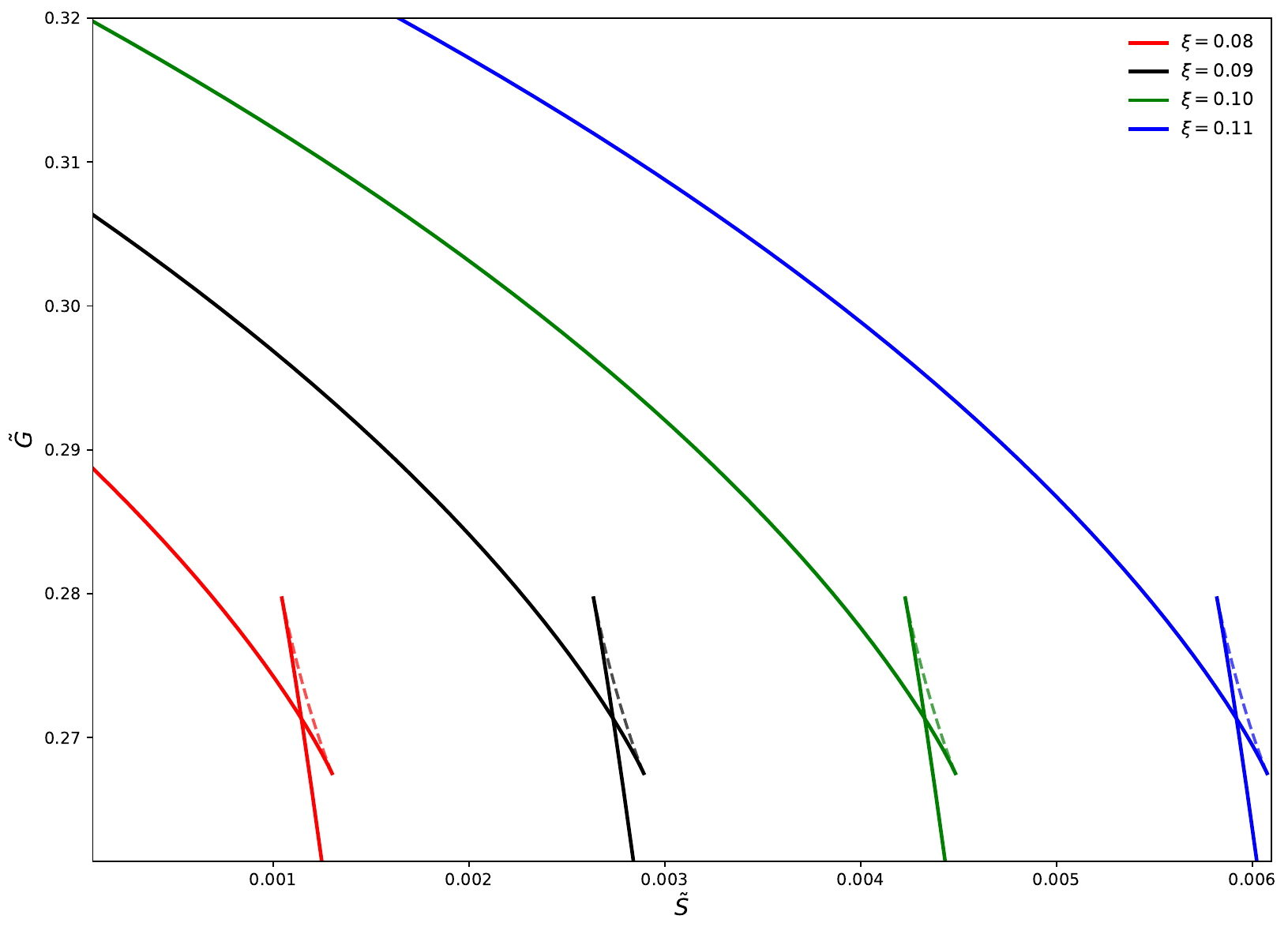}
\caption{$\tilde{G}$--$\tilde T_H$, $P=0.92P_c$.}
\label{fig:G_T_092}
\end{subfigure}
\hfill
\begin{subfigure}{0.32\textwidth}
\centering
\includegraphics[width=\linewidth]{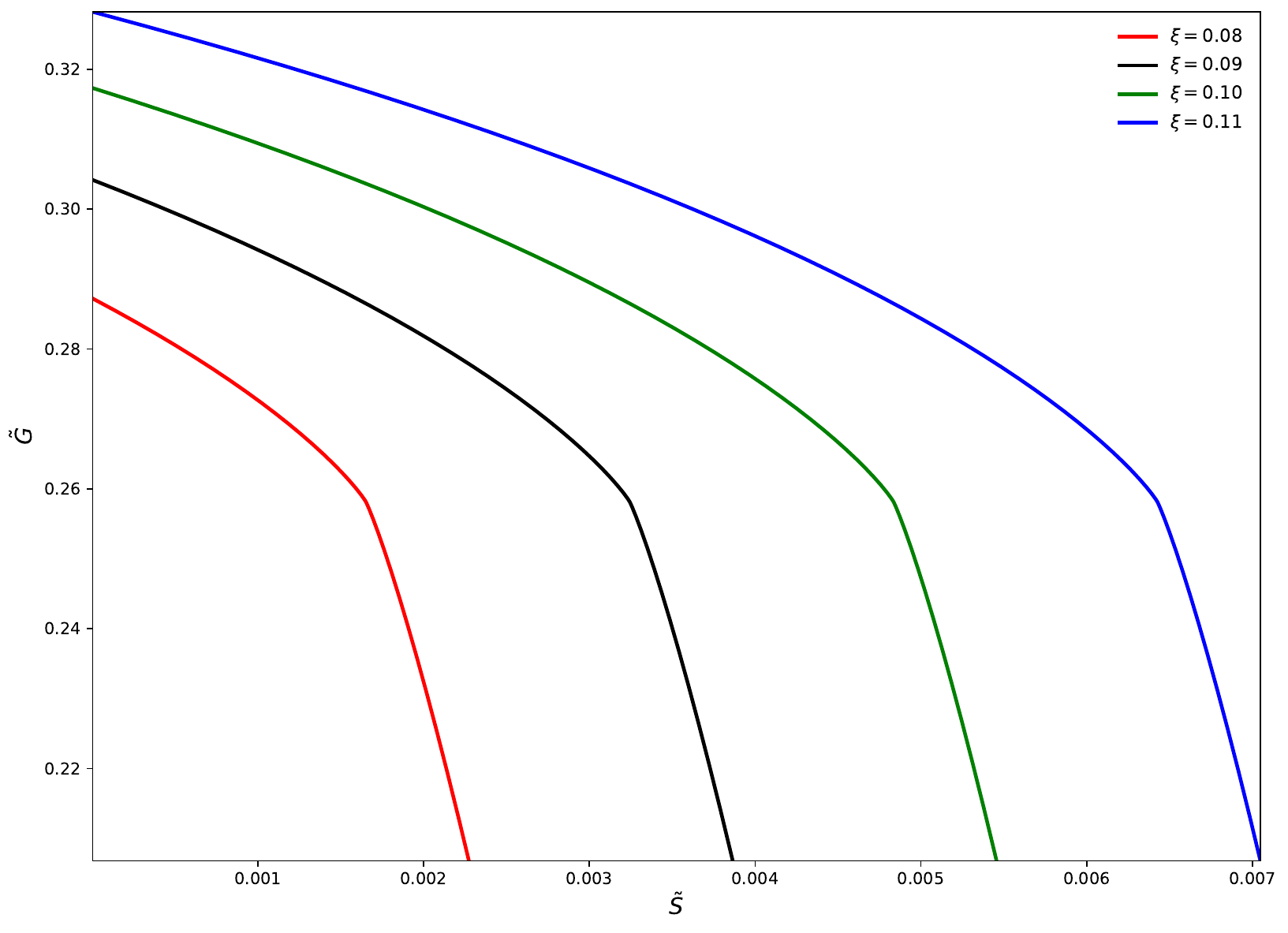}
\caption{$\tilde{G}$--$\tilde T_H$, $P=P_c$.}
\label{fig:G_T_100}
\end{subfigure}
\hfill
\begin{subfigure}{0.32\textwidth}
\centering
\includegraphics[width=\linewidth]{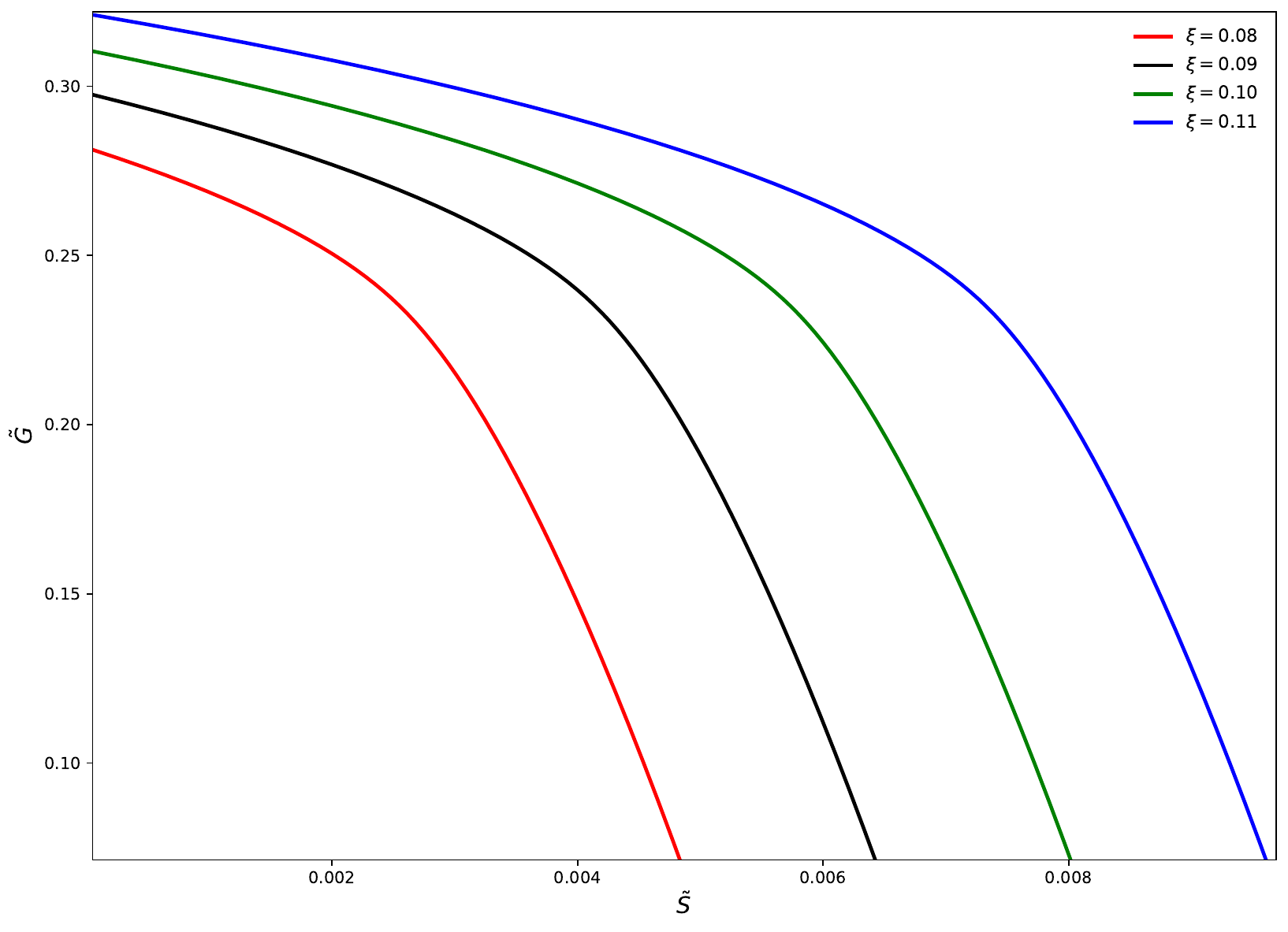}
\caption{$\tilde{G}$--$\tilde T_H$, $P=1.20P_c$.}
\label{fig:G_T_120}
\end{subfigure}

\vspace{0.3cm}

\begin{subfigure}{0.32\textwidth}
\centering
\includegraphics[width=\linewidth]{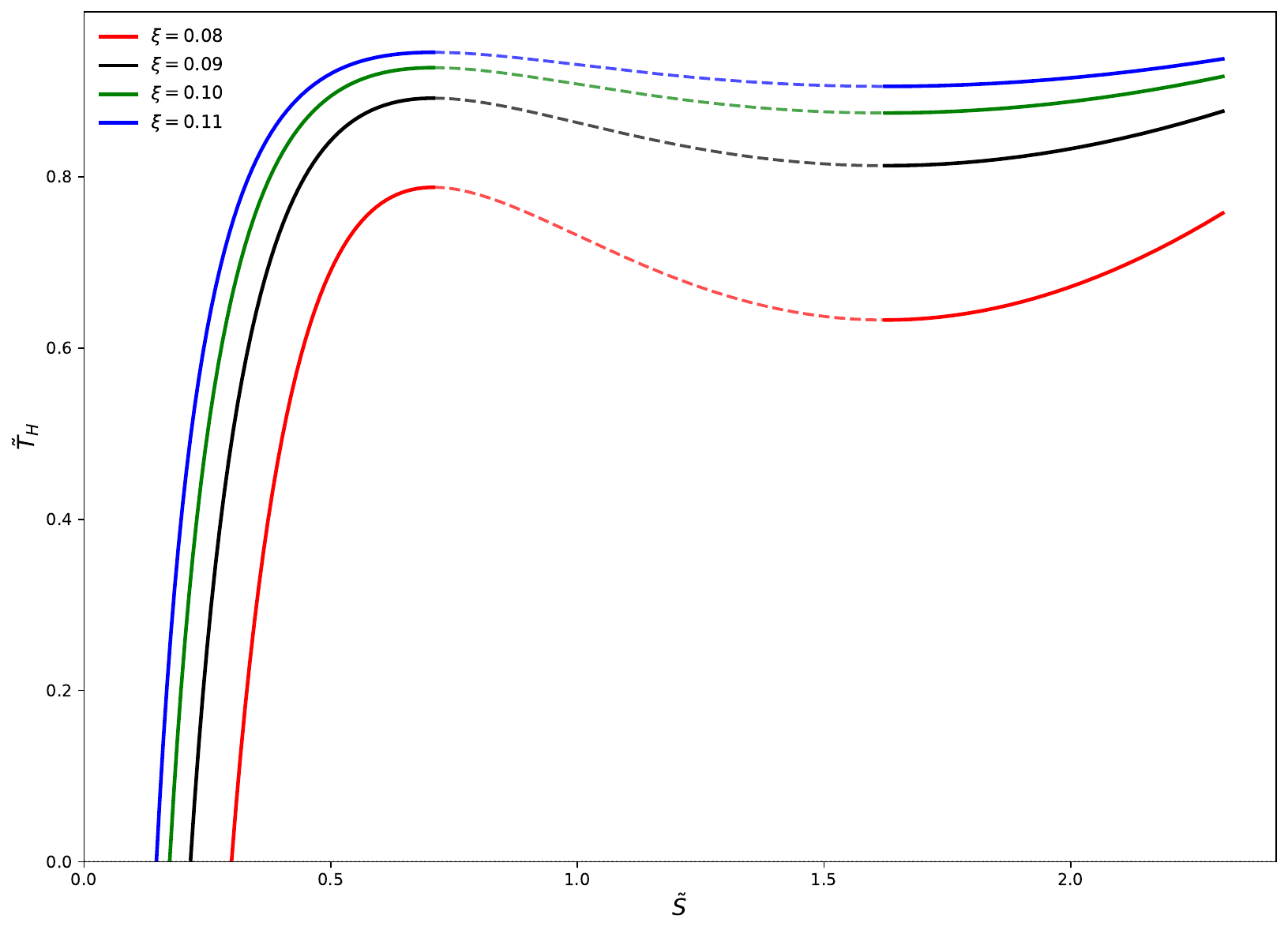}
\caption{$\tilde T_H$--$\tilde S$, $P=0.92P_c$.}
\label{fig:T_S_092}
\end{subfigure}
\hfill
\begin{subfigure}{0.32\textwidth}
\centering
\includegraphics[width=\linewidth]{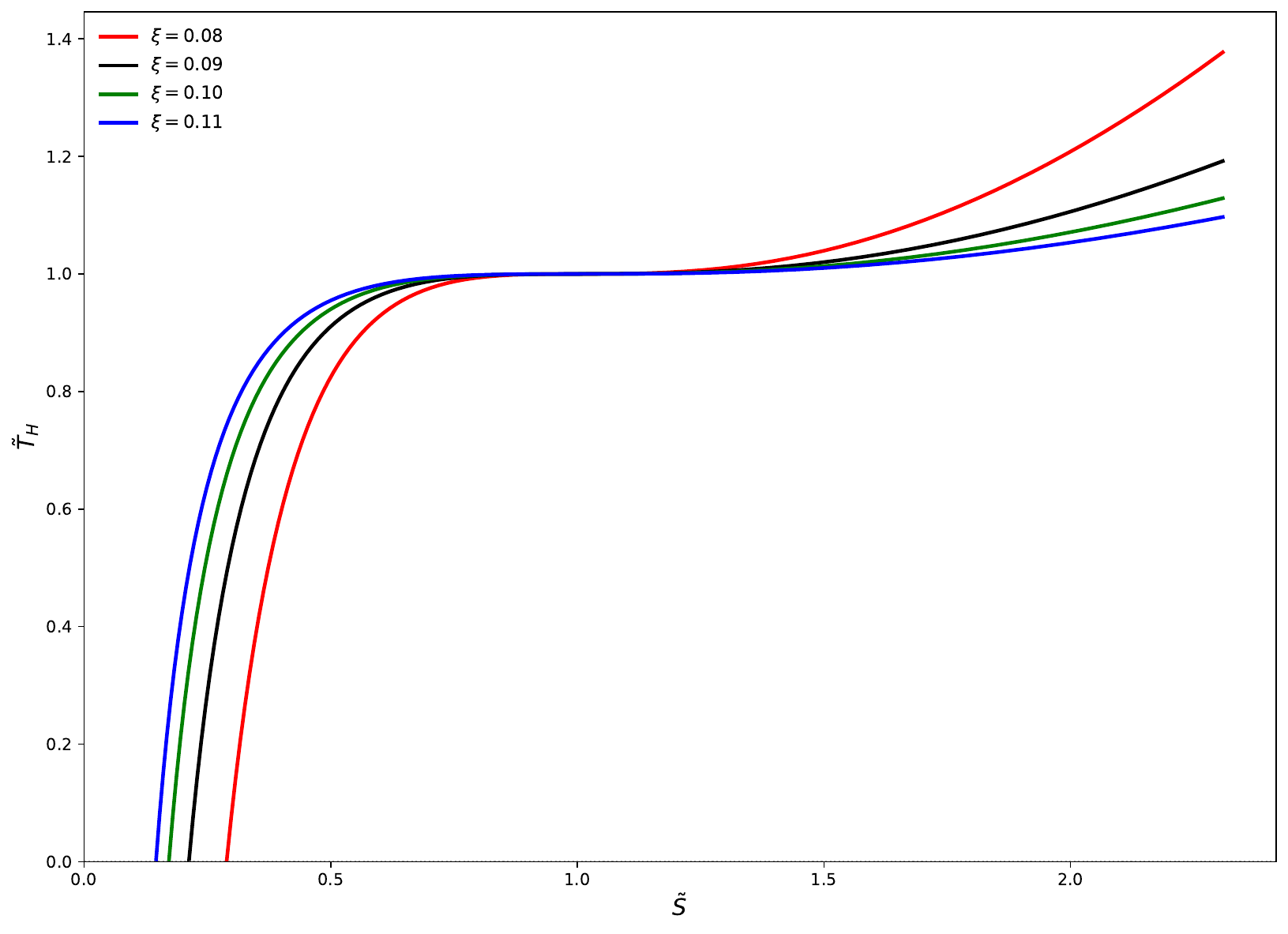}
\caption{$\tilde T_H$--$\tilde S$, $P=P_c$.}
\label{fig:T_S_100}
\end{subfigure}
\hfill
\begin{subfigure}{0.32\textwidth}
\centering
\includegraphics[width=\linewidth]{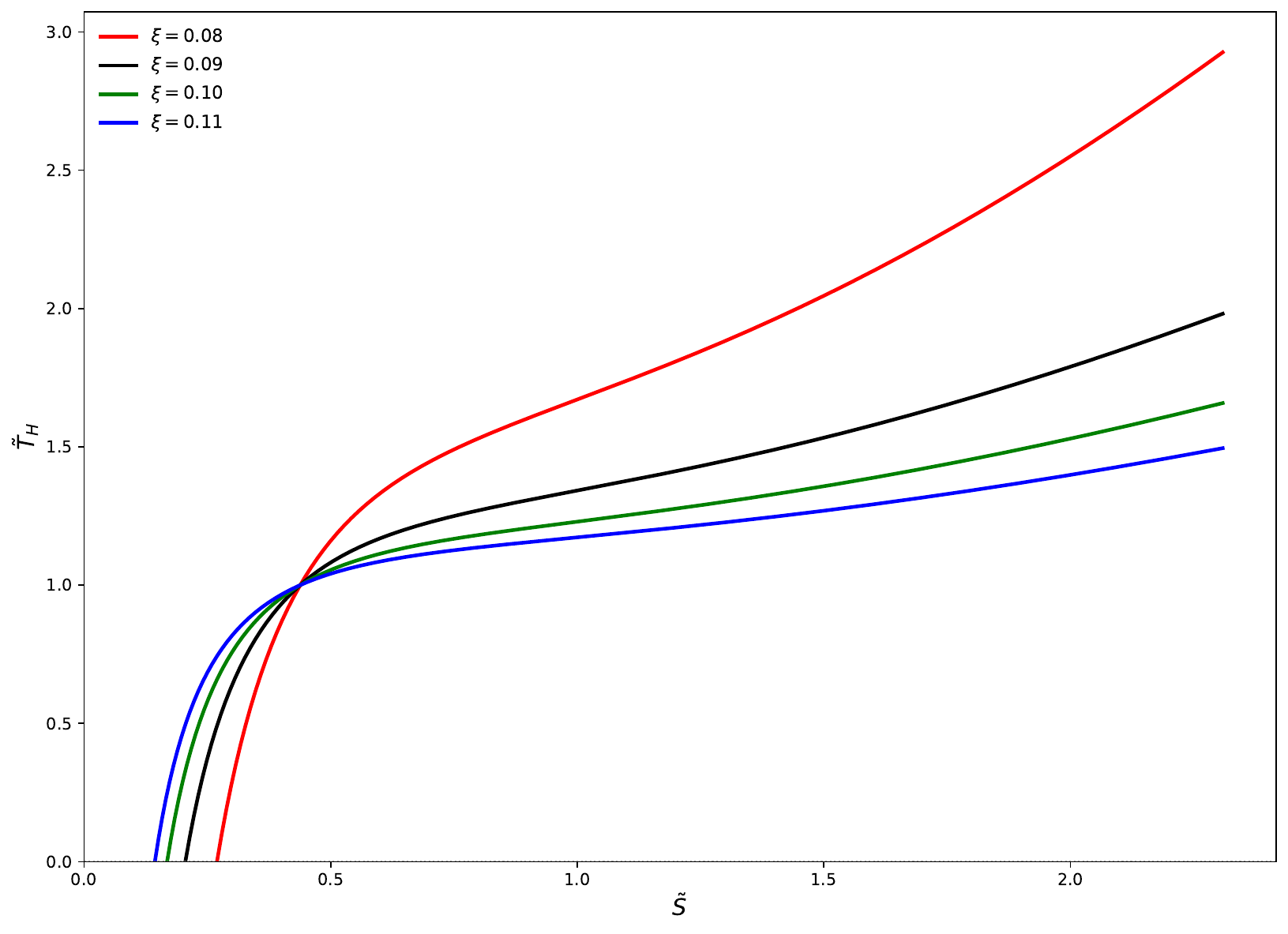}
\caption{$\tilde T_H$--$\tilde S$, $P=1.20P_c$.}
\label{fig:T_S_120}
\end{subfigure}

\vspace{0.3cm}

\begin{subfigure}{0.32\textwidth}
\centering
\includegraphics[width=\linewidth]{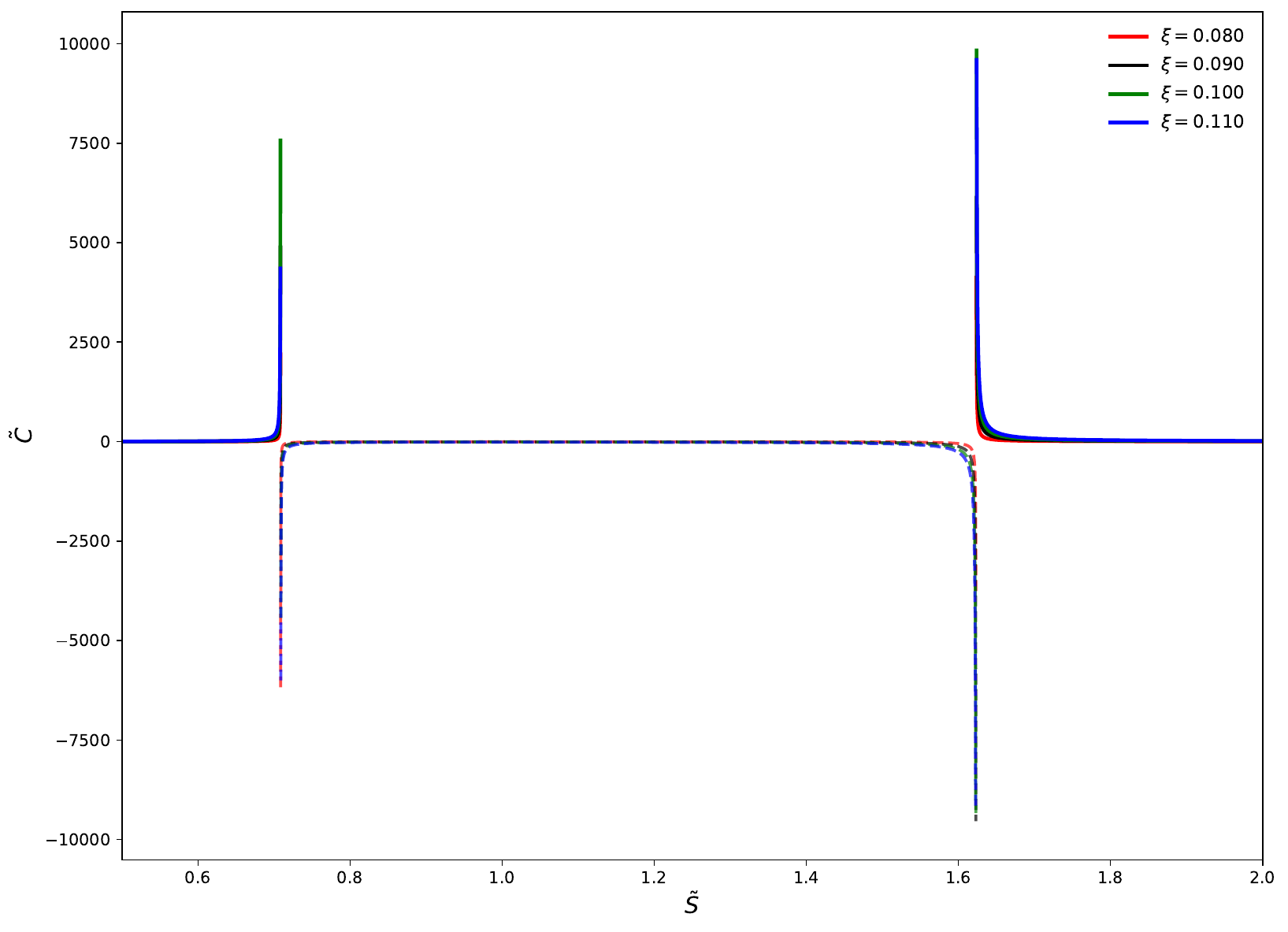}
\caption{$\tilde C$--$\tilde S$, $P=0.92P_c$.}
\label{fig:C_S_092}
\end{subfigure}
\hfill
\begin{subfigure}{0.32\textwidth}
\centering
\includegraphics[width=\linewidth]{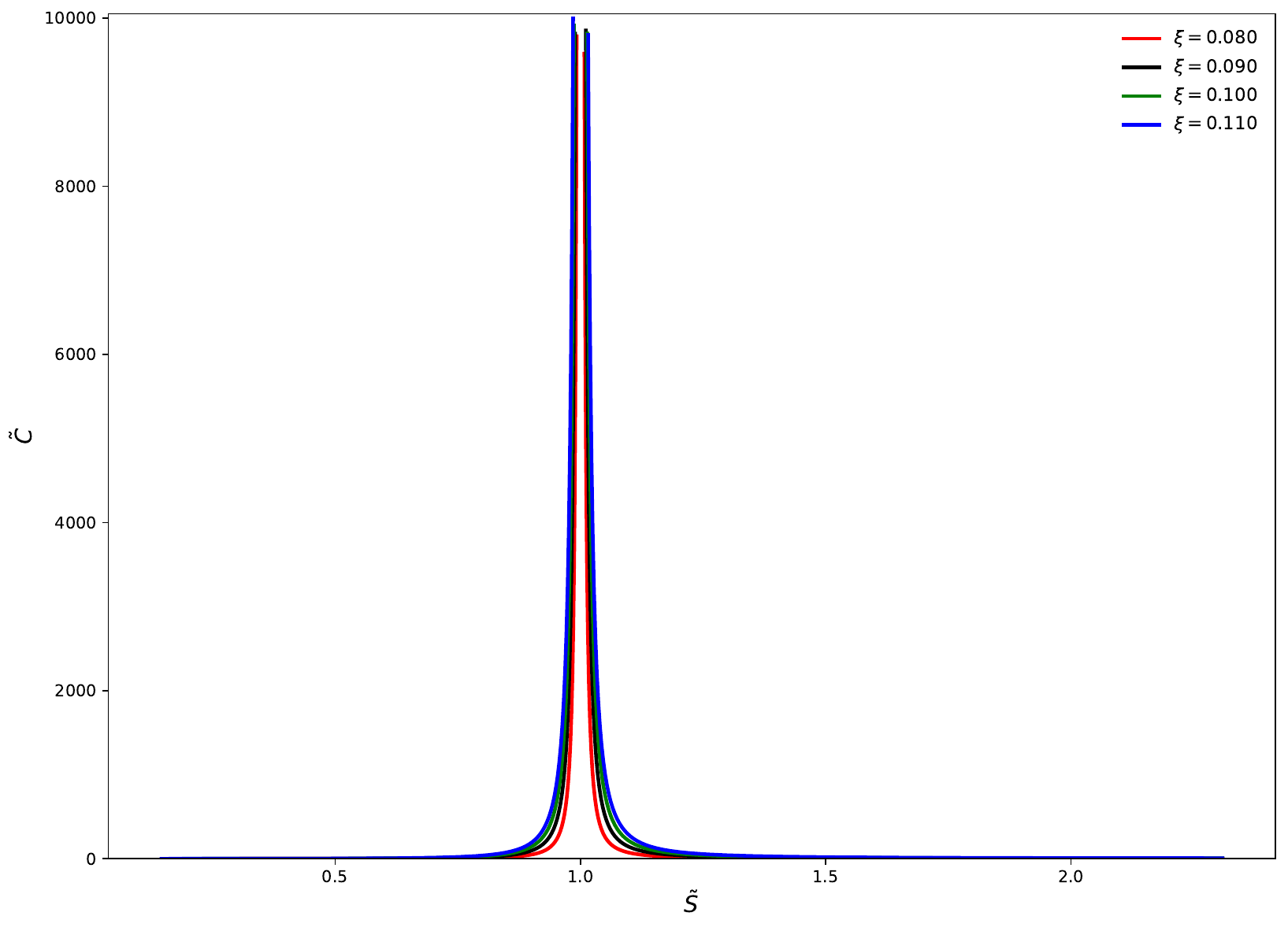}
\caption{$\tilde C$--$\tilde S$, $P=P_c$.}
\label{fig:C_S_100}
\end{subfigure}
\hfill
\begin{subfigure}{0.32\textwidth}
\centering
\includegraphics[width=\linewidth]{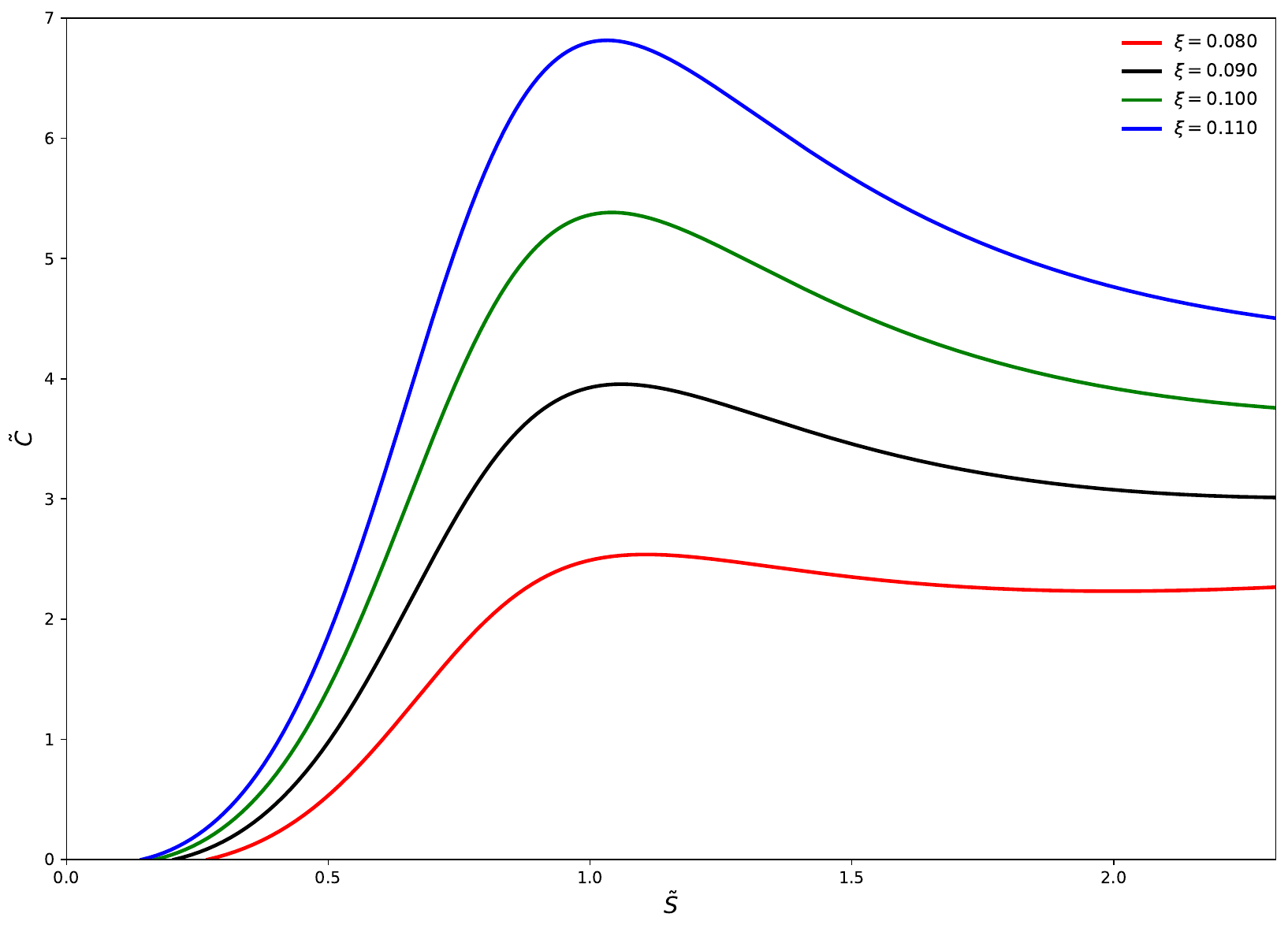}
\caption{$\tilde C$--$\tilde S$, $P=1.20P_c$.}
\label{fig:C_S_120}
\end{subfigure}

\caption{
Reduced thermodynamics of the generalized charged BTZ black hole.
}
\label{fig:thermo_nine_panels}
\end{figure}
For the numerical analysis, we use the following critical thermodynamic
values
\begin{equation}
S_c=43.31748,
\qquad
P_c=4.018230\times 10^{-4},
\qquad
T_c=1.651418\times 10^{-3},
\qquad
G_c=0.2581439 .
\label{eq:critical_values}
\end{equation}
Figure~\ref{fig:thermo_nine_panels} shows the thermodynamic response of
the generalized charged BTZ black hole for subcritical, critical and
supercritical pressures. The first row displays the reduced Gibbs free
energy as a function of the reduced Hawking temperature. The behavior of
$\tilde{G}$ allows us to identify the thermodynamically preferred branches:
configurations with lower Gibbs free energy are favored. As the pressure
increases from $P=0.92P_c$ to $P=1.20P_c$, the structure of the curves
changes, indicating that the pressure controls the possible transition
between different black-hole branches.

The second row shows the reduced Hawking temperature
$\tilde{T}_H=T_H/T_c$ as a function of the reduced entropy
$\tilde{S}=S/S_c$. Since the entropy is proportional to the horizon radius,
this plot describes how the temperature changes with the black-hole size.
The presence of extrema in $\tilde{T}_H(\tilde{S})$ signals changes in the
sign of $dT_H/dS$, which are directly related to changes of local
thermodynamic stability.

The third row presents the reduced heat capacity $\tilde{C}=C/C_c$.
Positive values of $\tilde{C}$ correspond to locally stable black-hole
branches, whereas negative values indicate thermodynamically unstable
configurations. The divergences of $\tilde{C}$ occur when $dT_H/dS=0$,
and therefore mark the boundaries between stable and unstable thermal
branches. In the present fixed-pressure Hawking-temperature description,
the parameter $\xi$ shifts the temperature and the free energy, while the
positions of the heat-capacity divergences are controlled by the zeros of
$dT_H/dS$.

\section{Discussions and conclusions}
\label{sec:conclusions}
In this work, we have investigated the optical and thermodynamic properties of a charged BTZ black hole surrounded by two effective matter sources, namely a quintessence-like anisotropic fluid and a cloud of strings. These additional sectors deform the standard charged BTZ geometry through linear and power-law corrections to the metric function, providing a simple framework for examining how exotic matter modifies black-hole physics in three-dimensional AdS spacetime.

The optical analysis reveals that the deformation introduced by the matter fields significantly changes the structure of null geodesics. In particular, the circular null-orbit condition,
$rf'(r)-2f(r)=0,$
admits an analytical solution in the special case $s=1$, where the orbit radius can be expressed in terms of the Lambert-$W$ function. Our stability analysis further shows that, throughout the physically relevant positive-density sector, these circular null orbits are stable. Consequently, they should be interpreted as trapped null trajectories rather than the unstable photon spheres that determine the boundary of a conventional black-hole shadow. This result illustrates that the existence of circular null orbits alone is not sufficient to guarantee the presence of a standard shadow, and that their stability properties are equally essential.

The absence of an unstable exterior photon sphere also influences the description of the energy emission process. In this case, the horizon circumference provides the relevant geometric absorption scale, and the emission spectrum is primarily governed by the horizon radius and the Hawking temperature. We find that both the quintessence-like matter and the cloud of strings enhance the emission rate through their effect on the horizon structure and surface gravity.

The thermodynamic behavior was examined by placing the black hole inside a finite cavity, thereby defining a canonical ensemble for the asymptotically AdS$_3$ spacetime. Using the Tolman redshift relation, we derived the local cavity temperature and analyzed the Gibbs free energy together with the heat capacity. The resulting phase structure demonstrates that the matter fields substantially modify the local thermal stability of the system, with the extrema of the temperature corresponding to transitions between stable and unstable thermodynamic branches.

Taken together, these results show that the combined presence of quintessence-like matter and a cloud of strings leads to qualitative modifications of both the optical and thermodynamic properties of charged BTZ black holes. Beyond shifting the location of the horizon and the thermodynamic quantities, these matter sectors alter the nature of null trapping and the corresponding thermal response, revealing features that are absent in the standard charged BTZ solution.

The present work can be extended in several directions. Including rotation would allow one to investigate the interplay between angular momentum, null geodesics, and cavity thermodynamics in matter-deformed BTZ spacetimes. It would also be interesting to study the corresponding quasinormal modes and holographic observables within the AdS$_3$/CFT$_2$ correspondence. Although the present analysis is performed in three dimensions, it provides a useful theoretical benchmark for understanding how additional matter sectors affect optical and thermodynamic observables, with potential implications for analogous four-dimensional black-hole models and their phenomenology. Finally, exploring possible Swampland constraints on these effective geometries may help clarify whether the matter sectors considered here admit a consistent embedding into quantum gravity.

\section*{Acknowledgments}
The National Center for Scientific and Technical (CNRST) funds the work of M. A. Rbah and S. Saoud 
Research  under the PhD-Associate Scholarship (PASS).
\bibliographystyle{elsarticle-num}
\bibliography{bib}

\end{document}